\documentclass[journal,twoside,web]{ieeecolor}
\usepackage{jsen}
\usepackage{cite}
\usepackage{amsmath,amssymb,amsfonts}
\usepackage{algorithmic}
\usepackage{graphicx}
\usepackage{textcomp}
\usepackage{wrapfig}

\usepackage{booktabs}
\usepackage[autostyle]{csquotes}
\usepackage{multirow}
\usepackage[draft]{hyperref}
\usepackage{scalerel}
\usepackage{tikz}
\usetikzlibrary{svg.path}
\usepackage{booktabs}

\definecolor{orcidlogocol}{HTML}{A6CE39}
\tikzset{
  orcidlogo/.pic={
    \fill[orcidlogocol] svg{M256,128c0,70.7-57.3,128-128,128C57.3,256,0,198.7,0,128C0,57.3,57.3,0,128,0C198.7,0,256,57.3,256,128z};
    \fill[white] svg{M86.3,186.2H70.9V79.1h15.4v48.4V186.2z}
                 svg{M108.9,79.1h41.6c39.6,0,57,28.3,57,53.6c0,27.5-21.5,53.6-56.8,53.6h-41.8V79.1z M124.3,172.4h24.5c34.9,0,42.9-26.5,42.9-39.7c0-21.5-13.7-39.7-43.7-39.7h-23.7V172.4z}
                 svg{M88.7,56.8c0,5.5-4.5,10.1-10.1,10.1c-5.6,0-10.1-4.6-10.1-10.1c0-5.6,4.5-10.1,10.1-10.1C84.2,46.7,88.7,51.3,88.7,56.8z};
  }
}

\newcommand\orcidi[1]{\href{https://orcid.org/#1}{\mbox{\scalerel*{
    \begin{tikzpicture}[yscale=-1,transform shape]
        \pic{orcidlogo};
    \end{tikzpicture}
}{|}}}}

\def\BibTeX{{\rm B\kern-.05em{\sc i\kern-.025em b}\kern-.08em
    T\kern-.1667em\lower.7ex\hbox{E}\kern-.125emX}}
\definecolor{abstractbg}{rgb}{0.89804,0.94510,0.83137}
\begin{document}
\title{Decoding EEG Rhythms During Action Observation, Motor Imagery, and Execution for Standing and Sitting}

\author{
    Rattanaphon~Chaisaen$^{\dagger \orcidi{0000-0003-1521-9956}}$, \IEEEmembership{Student member, IEEE}, Phairot~Autthasan$^{\dagger \orcidi{0000-0002-9566-8382}}$, \IEEEmembership{Student member, IEEE}, \vspace{-0.06in}\\
    Nopparada~Mingchinda, Pitshaporn~Leelaarporn$^{\orcidi{0000-0001-8755-875X}\,}$, Narin~Kunaseth, Suppakorn~Tammajarung, \vspace{-0.06in} \\
    Poramate~Manoonpong$^{\orcidi{0000-0002-4806-7576}\,}$,
    Subhas Chandra Mukhopadhyay$^{\orcidi{0000-0002-8600-5907}\,}$, \IEEEmembership{Fellow, IEEE}
    and \vspace{-0.06in} \\
    Theerawit~Wilaiprasitporn$^{\orcidi{0000-0003-4941-4354}\,}$, \IEEEmembership{Member, IEEE}
    \thanks{This work was supported by Robotics AI and Intelligent Solution Project, PTT Public Company Limited, Thailand Science Research and Innovation (SRI62W1501), and Thailand Research Fund and Office of the Higher Education Commission (MRG6180028).}
    \thanks{R. Chaisaen, P. Autthasan, N. Mingchinda, N. Kunaseth, S. Tammajarung, P. Leelaarporn and T. Wilaiprasitporn are with Bio-inspired Robotics and Neural Engineering (BRAIN) Lab, School of Information Science and Technology (IST), Vidyasirimedhi Institute of Science \& Technology (VISTEC), Rayong, Thailand. {\tt\small {corresponding author: theerawit.w  at vistec.ac.th}}}
    \thanks{P. Manoonpong is with BRAIN Lab, IST, VISTEC, Rayong, Thailand and Embodied AI \& Neurorobotics Lab, Centre for BioRobotics, The M{\ae}rsk Mc-Kinney M{\o}ller Institute, The University of Southern Denmark, Odense M, DK-5230, Denmark.}
    \thanks{S. C. Mukhopadhyay is with School of Engineering, Macquarie Uni-versity, Sydney, NSW 2109, Australia.}
    \thanks{\tt\small {$^{\dagger}$ equal contributions}}
    \thanks{Raw dataset, code examples, and other supporting materials are available on \url{https://github.com/IoBT-VISTEC/Decoding-EEG-during-AO-MI-ME}.}
}

\IEEEtitleabstractindextext{%
\fcolorbox{abstractbg}{abstractbg}{%
\begin{minipage}{\textwidth}%
\begin{wrapfigure}[12]{r}{3in}%
\includegraphics[width=2.85in]{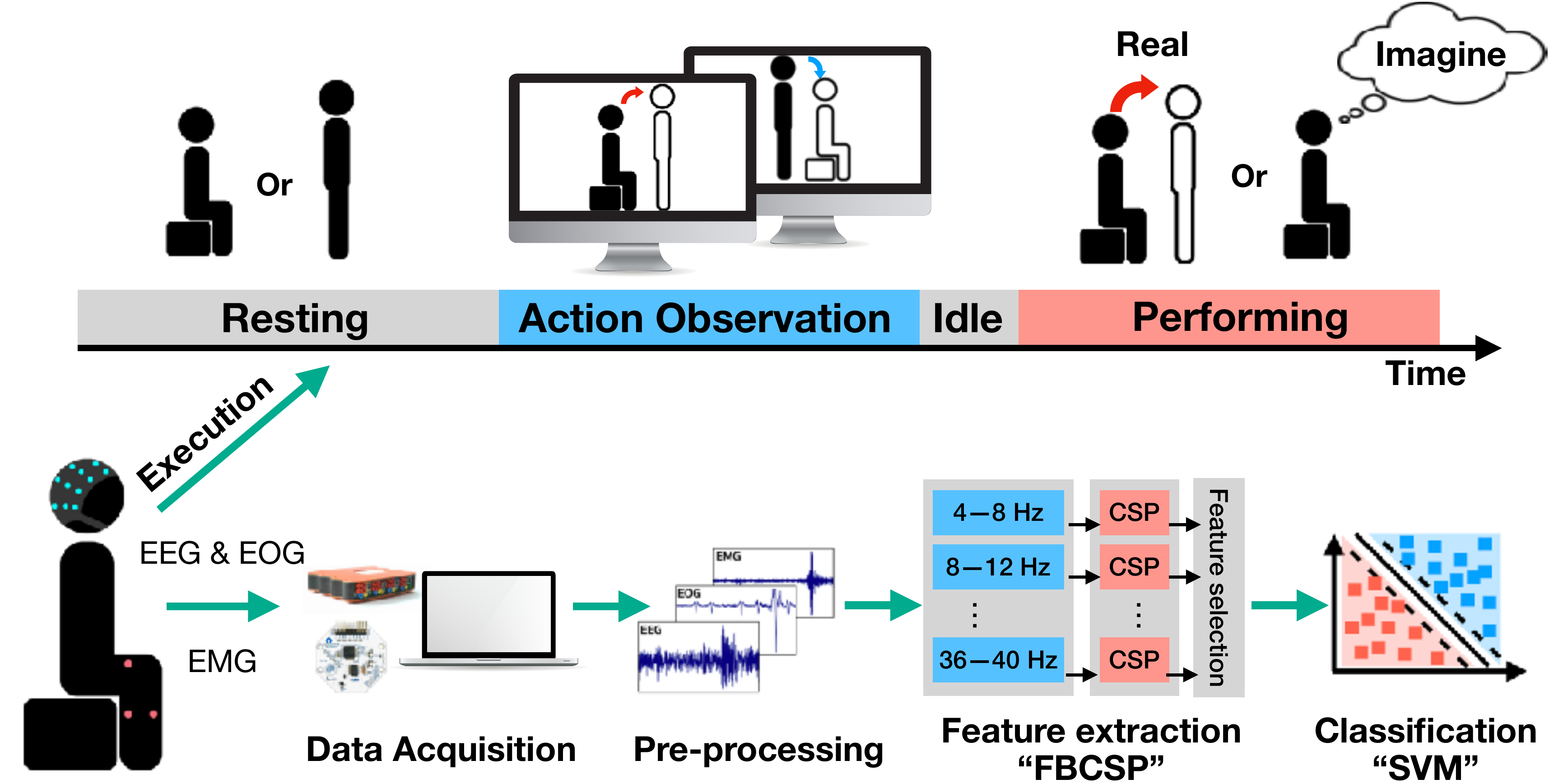}%
\end{wrapfigure}%
\begin{abstract}
Event-related desynchronization and synchronization (ERD/S) and movement-related cortical potential (MRCP) play an important role in brain-computer interfaces (BCI) for lower limb rehabilitation, particularly in standing and sitting. However, little is known about the differences in the cortical activation between standing and sitting, especially how the brain's intention modulates the pre-movement sensorimotor rhythm as they do for switching movements. In this study, we aim to investigate the decoding of continuous EEG rhythms during action observation (AO), motor imagery (MI), and motor execution (ME) for the actions of standing and sitting. We developed a behavioral task in which participants were instructed to perform both AO and MI/ME in regard to the transitioning actions of sit-to-stand and stand-to-sit. Our results demonstrated that the ERD was prominent during AO, whereas ERS was typical during MI at the alpha band across the sensorimotor area. A combination of the filter bank common spatial pattern (FBCSP) and support vector machine (SVM) for classification was used for both offline and classifier testing analyses. The offline analysis indicated the classification of AO and MI providing the highest mean accuracy at 82.73$\pm$2.54\% in the stand-to-sit transition. By applying the classifier testing analysis, we demonstrated the higher performance of decoding neural intentions from the MI paradigm in comparison to the ME paradigm. These observations led us to the promising aspect of using our developed tasks based on the integration of both AO and MI to build future exoskeleton-based rehabilitation systems.
\end{abstract}

\end{minipage}}}

\maketitle

\section{Introduction}
\label{sec:introduction}
\IEEEPARstart{T}{he} use of brain-computer interface (BCI) technology as a rehabilitation approach for motor disorders has become more extensive within the recent years. Within the past decade, there have been uses of BCI in a therapeutic setting, such as the use of motor imagery (MI) and virtual reality (VR) in post-stroke therapy \cite{ang2015randomized,ieee_sensor1,8823953}. The effectiveness of BCI technology in clinical settings has spanned to the development of exoskeleton for the rehabilitation of patients with multiple motor and motor-related disorders, such as upper limb exoskeleton\cite{frisoli2012new}. Indeed, BCI technology has been found to be an effective rehabilitation approach to motor related complications as a result of stroke, for example, with an increase in the upper limb strength as measured by the Fugl-Meyer Motor Assessment (FMMA) after the implementation of MI-based BCI \cite{ang2015brain,foong2019assessment}. In order for the BCI technology to be effective, it is essential that the users are able to control the exoskeleton system via methods such as biofeedback from electroencephalography (EEG), electromyography (EMG), and electrooculography (EOG)\cite{kwak2015lower,7562544,7459401}.

Event-related desynchronisation/synchronisation (ERD/S) are cortical rhythms characterized by the mu (8--13 Hz) and beta (14--30 Hz) neural activity patterns \cite{pfurtscheller1999event, kitahara2017target}. As ERD/S are prominent during MI of limb movements, ERD/S-based BCI has shown potentials for the rehabilitation of motor disorders \cite{kitahara2017target, ieee_sensor2}. Supporting this notion of ERD in motor preparation and inhibition is the study on experimental participants in a unilateral wrist extension task based on visual cues, in which mu ERD showed stronger contra-lateralization features with movement intention and execution in the sensorimotor cortices \cite{li2018combining, lee2019eeg} whereas ERS was found prominently in the ipsilateral hemisphere \cite{cho2017eeg}. In order to implement ERD/S-based BCI in exoskeleton, EMG is often used to modulate the gait pattern of the exoskeleton, whilst algorithms such as common spatial patterns (CSP) has been used to decode the MI task done by a participant \cite{li2019hybrid}. However, another factor that must be taken into account is the motor planning, which involves the intention of a person prior to the execution of a movement. One way in which the rehabilitation via BCI can be achieved is by altering the neural activities of a person using methods such as modulating pre-movement sensorimotor rhythm (SMR) \cite{McFarland_2015}. In rehabilitative settings, SMR can be altered via instructions, a process known as learned modulation \cite{norman2016movement}. There is evidence supporting the effectiveness of learned modulation on motor performance after training with EEG-based feedback\cite{meng2016noninvasive, sarasola2017hybrid}. Specifically, the decrease in the amplitude of pre-movement SMR correlated with a more accurate performance in a target matching task, implicating that ERD will support motor functions due to the correlation between ERD and motor accuracy \cite{McFarland_2015}.

Movement-related cortical potentials (MRCPs) are spontaneous potentials that are generated during a person-generated planning and motor execution (ME), which can be an actual movement or imaginary \cite{jeong2020decoding, article_used}. MRCP generally consists of two main parts called a Bereitschaftspotential (BP), or readiness potential (RP), and a movement-monitoring potential (MMP). In addition to ERD/S, MRCP can be used to decode motor intention and planning\cite{jeong2020decoding}. One aim of the current study is to evaluate the competencies of ERD/S and MRCP in terms of their reflections of movement intentions, specifically regarding the contributions of the current research design on future research in rehabilitative exoskeleton systems. In healthy participants, there are no effects of subject training on MRCP-based BCI technology\cite{jochumsen2018effect}. In motor rehabilitation, MRCP is thought to underlie neuroplasticity, which can be implemented faster when using BCI systems with high signal-to-noise ratio and lowered calibration time \cite{colamarino2019adaptive, lin2016discriminative}.

Nevertheless, only a small amount of studies has been conducted on how the brain mediates different complex gait movements such as running and walking \cite{goh2018spatio}. As there is less work on the complexity of shifting from sitting to standing and vice versa \cite{sit_stand_paper}, the current study aimed to combine action observation (AO) and MI as a potential rehabilitation strategy for lower limbs dysfunction. To assess the roles of AO, continuous EEG rhythms were collected throughout the entire experimental procedure, which included AO, MI, and resting (R) states. Participants were instructed to perform both actions of AO and MI in regard to standing and sitting, which were alternated between the actions of sit-to-stand and stand-to-sit. Due to the increase in the alpha and beta patterns as a result of \enquote{push-off} (heel striking) actions, we expected to see the differences in the cortical activation between standing and sitting in the sense that the act of transitioning from sitting to standing would result in the act of push-off \cite{nordin2019faster}.

\begin{figure}
    \centering
    \includegraphics[width=1.00\columnwidth]{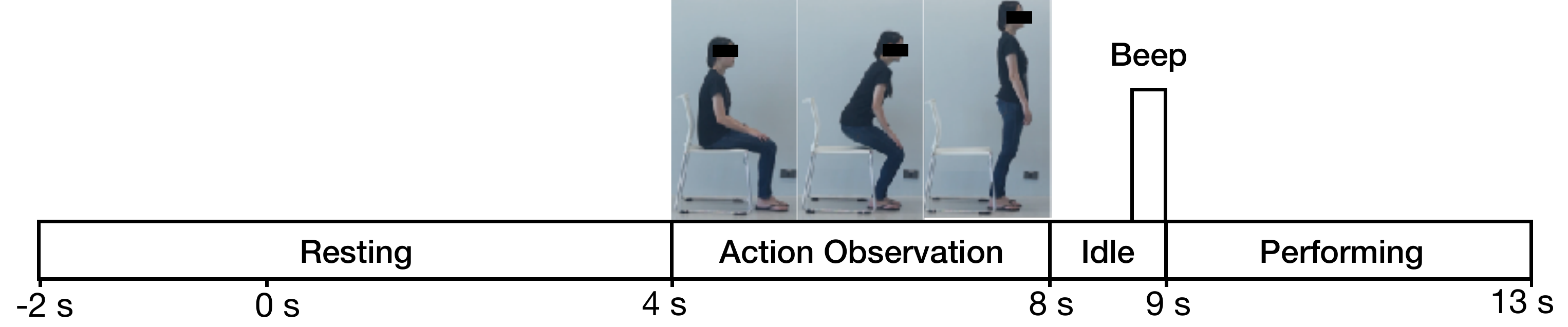}
    \caption{Timeline of each experimental trial. The four states displayed include resting/R (0--4 s), action observation/AO (4--8 s), idle (8--9 s), and task performing, either MI or ME (9--13 s).}
    \label{timeline}
\end{figure}

There are two major contributions of the current study:
\begin{enumerate}
   \item The current study aims to explore the lower limb motor functions using action observation (AO), motor imagery (MI), and motor execution (ME) together before distinguishing between individual EEG correlates, with each state showing different cortical activation patterns. Specifically, we looked at the EEG potentials during resting, sit-to-stand, and stand-to-sit. In each trial, a video of a person performing the actions was shown. The design is expected to facilitate the future exoskeleton-based rehabilitation that integrates both AO and MI. This is further discussed in \textit{Discussion A}.
   \item Using the state-of-the-art machine learning algorithms and classifier testing scheme, our classification approaches are shown to distinguish between the resting (R) versus AO and AO versus task performance (MI or ME). Although multivariate pattern analysis (MVPA) was proposed as an analysis method for multiple brain activations across participants, the order of participants entered was found to be sensitive \cite{al2020hyperalignment}. With classifier testing scheme and two classifications, we enabled the practical assessment of classifier performance in this study. Further discussion regarding the algorithms and EEG classifications can be found in \textit{Discussion C}. 
\end{enumerate}
    
From our findings, we aimed to contribute to a smoother interface between user and the exoskeleton system, which is the main challenge of implementing rehabilitative exoskeleton technology \cite{young2016state}.

\section{Methods}
\subsection{Participants}
The recruited participants comprised 8 healthy individuals (3 males, 5 females; 20--29 years old) with no history of neurological disorder, lower limb pathology, or gait abnormalities. All participants gave their informed consent prior to the experimental procedure following the Helsinki Declaration of 1975 (as revised in 2000). The study was approved by Rayong Hospital Research Ethics Committee (RYH REC No.E009/2562), Thailand.
\newline

\begin{figure}
    \centering
    \includegraphics[width=0.7\columnwidth]{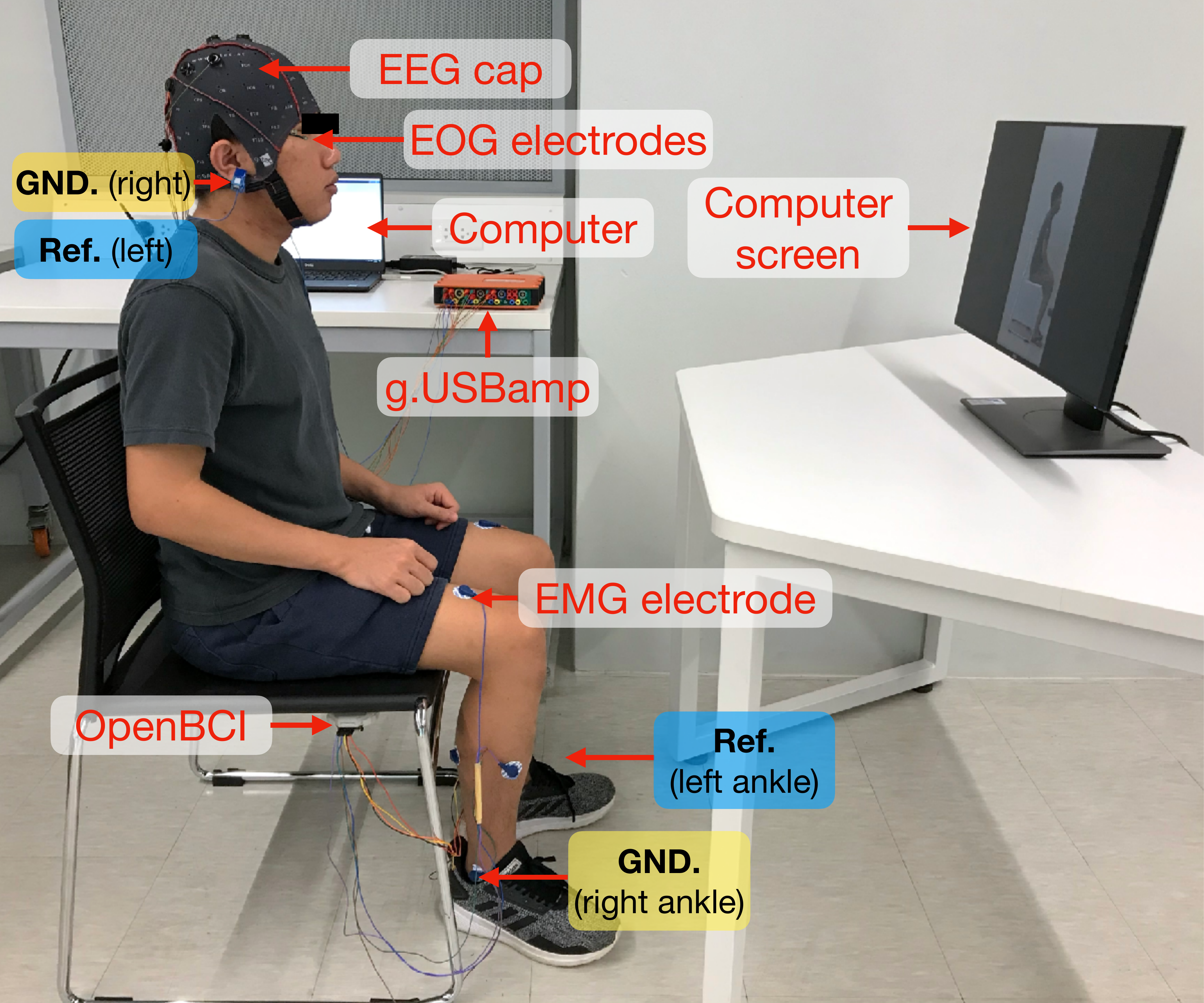}
    \caption{The sensing system set up for EEG, EMG, and EOG data acquisition.}
    \label{sensor_setup}
\end{figure}

\subsection{Experimental Protocol}

To investigate the feasibility of decoding the MI signals (including ERD/S) and MRCPs during the intended movement executions with continuous EEG recordings, the entire experimental procedure composed of two sessions: MI and ME. Each session consisted of 3 runs (5 trials each), incorporating a total of 30 trials. The protocol began with a sitting posture, followed by 5 repeated trials of sit-to-stand and stand-to-sit tasks alternatively. \autoref{timeline} displays the sequence of four states in each trial: R, AO, idle, and task performing (MI or ME). During the R state, a black screen was displayed on the monitor for 6 seconds (s). The participants were instructed to remain relaxed and motionless. To avoid the ambiguity of the instructions, a video stimulus showing either the sit-to-stand or stand-to-sit video task, lasted for 4 to 5 s, was presented to guide the participants in the AO state. The participants were instructed to perform the tasks of both sessions immediately after the audio cue. In the ME session, the participants were to complete a succession of self-paced voluntary movements after the audio cue. In the MI session, the participants were to commence the imagining of the assigned motion after the audio cue. During MI, the motor initiation onset can be generally obtained from an audio cue or visual cue, whereas during ME, the motor initiation onset from EMG signals.

\subsection{Data Acquisition}
 The sensing system was set up to record the EEG, EOG, and EMG signals simultaneously throughout the experiment, as depicted in \autoref{sensor_setup}. A biosignal amplifier (g.USBamp RESEARCH, g.tec, Austria) was used to acquire EEG and EOG signals. The EEG signals were obtained using 11 passive electrodes, positioned according to the 10-20 international system at the following placements: $FCz$, $C3$, $Cz$, $C4$, $CP3$, $CPz$, $CP4$, $P3$, $Pz$, $P4$, and $POz$, with the reference and ground electrodes placed on the left and the right earlobes, respectively. EOG signals were acquired from 2 passive electrodes positioned under and next to the outer canthus of the right eye. The impedance of both EEG and EOG signals was maintained at below 10 k$\Omega$ throughout the experiment and the sampling rate was set to 1200 Hz. Moreover, an open-source, low-cost, and consumer-grade biosignal amplifier, namely \textit{OpenBCI}, was used to recorded EMG signals to identify the onset of the movement. The device is developed based on the Analog Front-End (AFE) ADS1299 (Texas Instruments, USA) \cite{s18113721}. Currently, many studies published in high reputation journals have focused on the usability of OpenBCI device in a variety of BCI applications \cite{max_paper, kim_paper, tan_paper}. 6 electrodes were placed on rectus femoris (RF), tibialis anticus (TA), and gastrocnemius lateralis (GL) of two lower limbs with sampling frequency of 250 Hz. The reference and ground electrodes were placed at the left and the right ankles, respectively.

\subsection{EEG Pre-processing}

\begin{figure}[t]
    \centering
    \includegraphics[width=1.0\columnwidth]{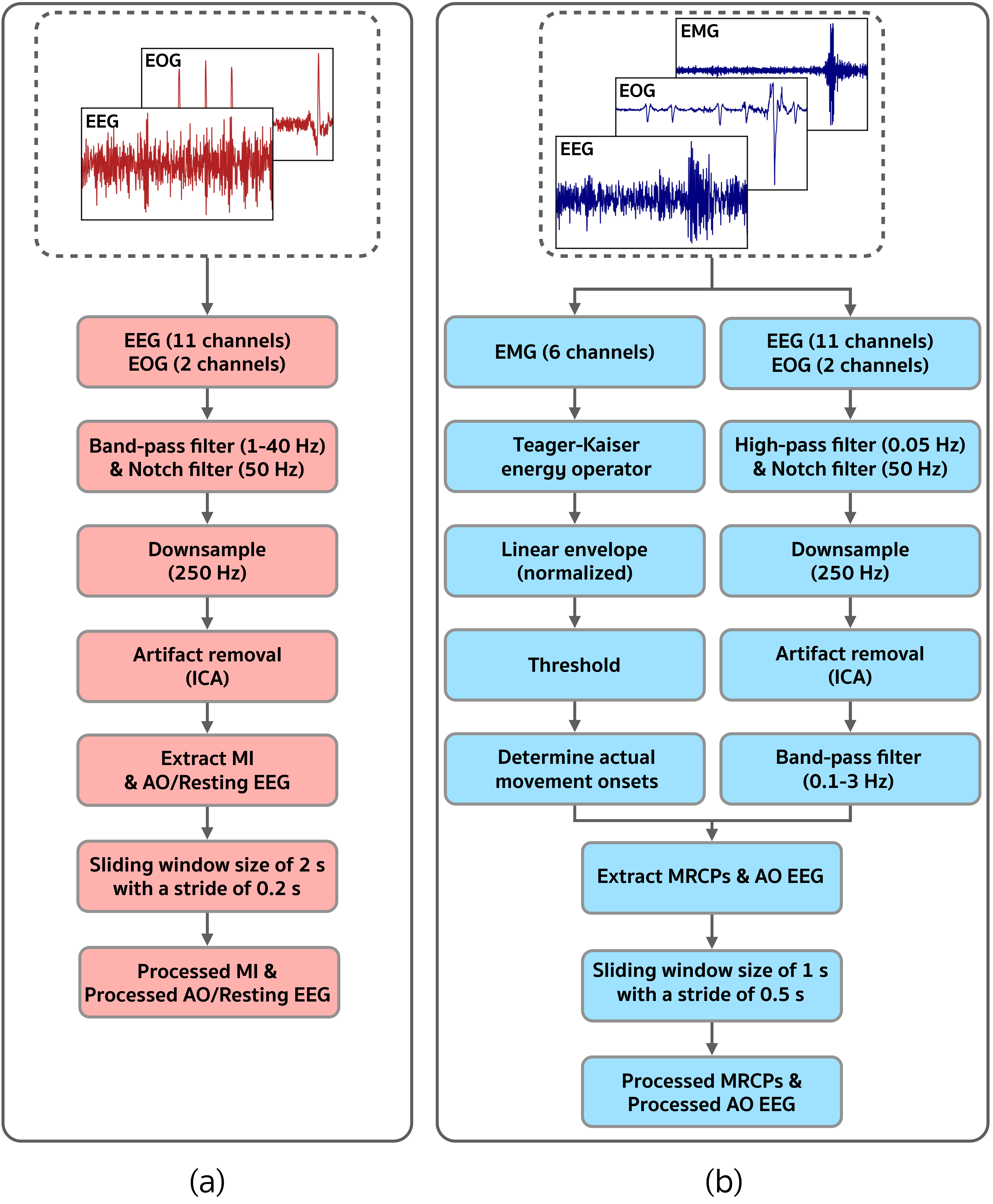}
    \caption{Overview of the EEG, EOG, and EMG data pre-processing. (a) exhibits the procedure of MI signal pre-processing on the EEG and EOG data from the MI. (b) illustrates the pre-processing steps to extract MRCPs from the ME.}
    \label{preprocess_eeg}
\end{figure}

The offline signal processing was performed using MNE-python package (version 0.20.0) \cite{MNE}. The pre-processing process was divided into two main steps: EEG-based MI signal and EEG-based MRCP during both MI and ME. \autoref{preprocess_eeg} illustrates the course of EEG, EOG, and EMG data processing.

\textit{Motor Imagery}: The notch filter was set at 50 Hz to reduce the electrical noises. The recorded EEG and EOG signals were band-pass filtered between 1--40 Hz, using 2\textsuperscript{nd} order non-causal Butterworth filter. Both signals were down-sampled to 250 Hz. An eye movement-related artifact correction-based on independent component analysis (ICA) \cite{fastICA} was applied to the EOG signals for the identification of artifact components which were removed from the EEG data. The EEG signals were segmented in epochs locked to the onset of each class (R, AO, and MI) for 4 s, as shown in \autoref{timeline}, followed by the pre-processing using a 2 s sliding window with a 0.2 s shift. The processed data for each class from each participant contained a collection of trials$\times$windows$\times$channels$\times$time points (15$\times$11$\times$11$\times$500).

\textit{Movement-related Cortical Potentials}: A threshold-based method generally plays a significant role in extracting the actual movement onset detected by the EMG \cite{8410007}. In this study, we employed the threshold-based method to determine the actual movement onset of each sit-to-stand/stand-to-sit transition. The Teager-Kaiser energy operator (TKEO) \cite{tkeo2} was firstly applied to each EMG channel to enhance the signal-to-noise ratio for the onset detection. The signals were then band-pass filtered between 15--124 Hz (2\textsuperscript{nd} order non-causal Butterworth filter), rectified using the absolute value and low-pass filtered at 3 Hz (2\textsuperscript{nd} order non-causal Butterworth filter) to compute the linear envelope. A time window of 2 s before the audio cue was selected as the reference signal as no explicit movement should be occurred in this time interval. The linear envelope of the signals was applied to calculate the threshold ($T$), which was set as $T = \mu + h * \sigma$, where $\mu$ and $\sigma$ were the mean and standard deviation (SD) of the reference signal. Moreover, $h$ was varied from 3 to 20 where the highest classification accuracy was selected from. $h$ = 10 was used for the calculation of $T$. Owing to the concerned related to the fallibility of identifying the movement onset, the onset was determined by considering the number of consecutive samples ($E$) where the EMG envelopes exceeded the $T$. We empirically defined $E$ as 5. Therefore, the first time point was marked as the actual movement onset, when more than $E$ (5) consecutive samples exceeded the $T$.

The MRCP signals were extracted from the EEG signals recorded during the ME. The EEG signals were then high-pass filtered at 0.05 Hz (2\textsuperscript{nd} order non-causal Butterworth filter). The notch filter frequency rate was defined at 50 Hz to filter out the electrical noises. Next, the EEG signals were down-sampled from 1200 to 250 Hz. The eye movement-related artifacts were removed, using the same ICA as the EEG-MI pre-processing protocol. The time-locked EEG and EMG signals were segmented into pre-movement and resting periods, based on the actual movement onset from the EMG signals. Each pre-movement epoch comprised the EEG signals from -1.5 to 1 s, which were identified as the \enquote{MRCPs} class. Each resting epoch consisted of the EEG signals in AO state from 4 to 6.5 s, defined as the \enquote{AO} class. After the MRCPs extraction was completed, the data from each class were converted into a sequence of sub-samples or sliding windows with a 1 s sliding window and a step of 0.5 s. The processed MRCPs and AO data from each participant were formed in a dimension of trials$\times$windows$\times$channels$\times$time points (15$\times$4$\times$11$\times$250).

\subsection{Qualitative Analysis}

Time-frequency analysis was utilized using EEGLAB toolbox (version 2019.0) \cite{eeglab} to visualize sit-to-stand and stand-to-sit executions during the MI session after performing ICA method in the aforementioned pre-processing. Event-Related Spectral Perturbations (ERSP) method \cite{ERSP} was performed for the frequency ranges from 4 to 40 Hz for all channels to compute the power spectra by applying the Morlet wavelets transform with incremental cycles (3-cycle wavelet at the lowest frequency, up to 15 Hz at the highest), resulting in 200 time points. The baseline reference was then taken from -1 to 0 s at the beginning of the R state. The average of the spectral power changes was calculated at each time while normalized by the baseline spectra. The significance of deviations from the baseline was analyzed using bootstrap method ($p=0.05$).

To exhibit the qualitative result of MRCPs, all of 2.5 s from the extracted MRCPs (15 trials executed by each participant) were considered. The signals were then re-referenced using current source density (CSD) with spherical spline interpolations to enhance the unsatisfactory spatial resolution of EEG data \cite{7979603}. The CSD was applied to the extracted MRCPs from all 11 EEG channels to remove the overall background activity. Subsequently, the grand average MRCP waveform was generated for each sit-to-stand/stand-to-sit transition using the average value of every trial across the 8 participants.

\subsection{Offline Analysis}

\begin{figure*}
    \centering
    \includegraphics[width=2.0\columnwidth]{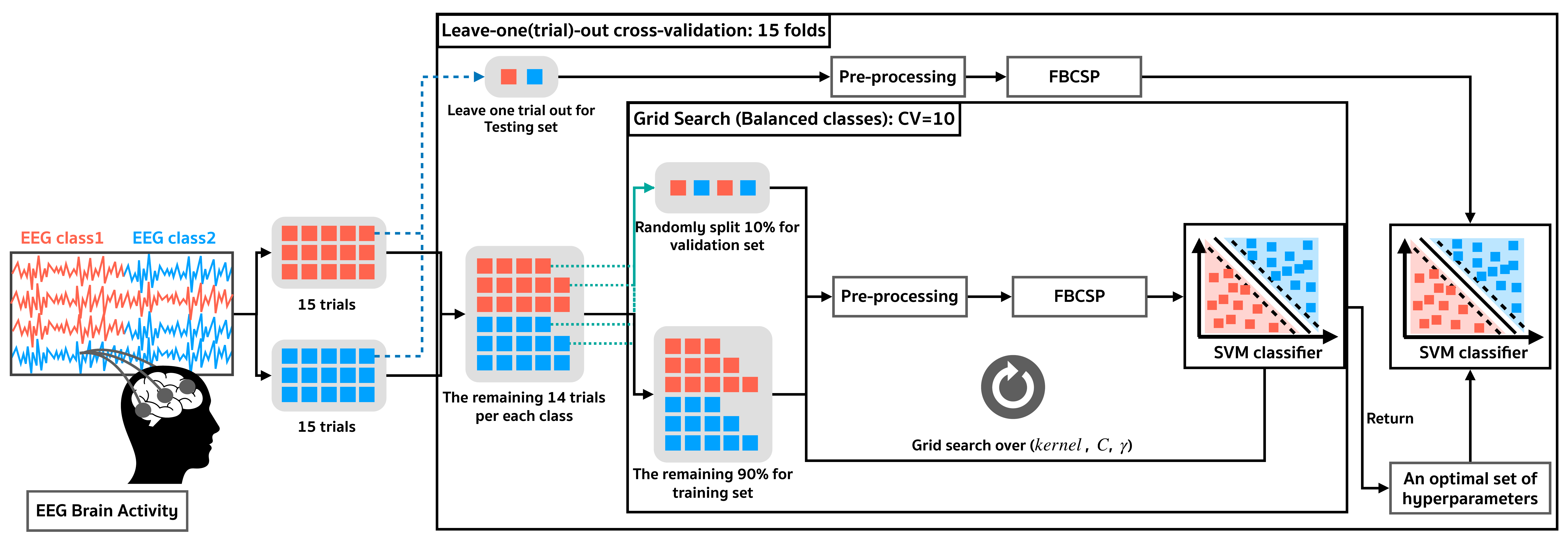}
    \caption{Architecture of leave-one(trial)-out cross-validation (LOOCV) with the grid search algorithm for the binary classification models. LOOCV was performed independently subject by subject.}
    \label{GridSearch}
\end{figure*}

\begin{figure}[t]
    \centering
    \includegraphics[width=1\columnwidth]{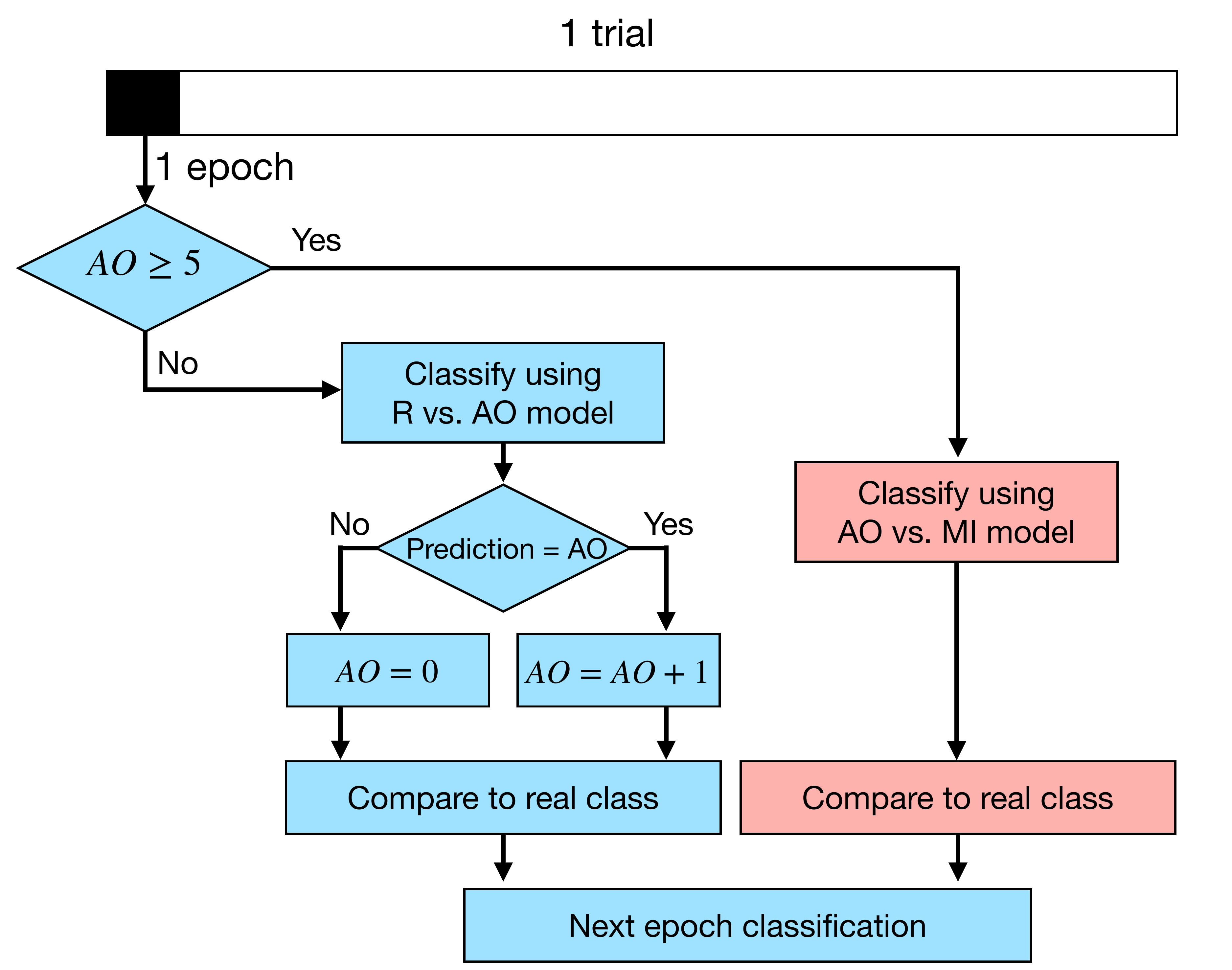}
    \caption{Flowchart of classifier testing analysis used in the MI session. The grid search algorithm was applied to assist in determining the action observation versus motor imagery (AO vs.MI) classification model when the action observation was produced after 5 consecutive detections ($AO\geq5$).}
    \label{pseudo_online_flowchart}
\end{figure}

To demonstrate the possibility of decoding the MI signals and MRCPs, the binary classification tasks on both signals were designed based on the exoskeleton system with the ability to identify and control each exact movement (standing or sitting). Thus, each sit-to-stand/stand-to-sit transition was conducted using the classification tasks separately. In the MI session, two classification tasks (R versus AO and AO versus MI) for neural decoding of the standing and sitting movements were conducted. In the ME session, only one classification task (AO versus MRCPs) was performed.

Subject independent classification tasks were implemented with leave-one(trial)-out cross-validation (LOOCV) on 15 trials (15 folds) using Scikit-learn \cite{scikit-learn}, as exhibited in \autoref{GridSearch}. Each fold composed of 14 trials as the training set and the remaining trial as the testing set. During the training session, signal pre-processing was firstly performed on the training set, as depicted in \autoref{preprocess_eeg}. Furthermore, the spatial features were extracted from the sub-sampled training set using the filter bank common spatial pattern (FBCSP), producing the feature vectors for the classification task. Importantly, the FBCSP performs generally well in MI classification tasks \cite{8310961}. The FBCSP was introduced as an extension of the common spatial pattern (CSP) to autonomously select the discriminated EEG features from multiple filter banks. In this study, 9 filter bank, arrays of band-pass filters, using 2\textsuperscript{nd} order non-causal Butterworth filter with a bandwidth of 4 Hz from 4 to 40 Hz (4--8 Hz, 8--12 Hz, ..., 36--40 Hz) were created for the MI. 6 filter banks were built with a bandwidth of 0.4 Hz for the 1\textsuperscript{st} band and a bandwidth of 0.5 Hz for the other bands from 0.1--3 Hz (0.1--0.5 Hz, 0.5--1 Hz, ..., 2.5--3 Hz) for the ME.

Subsequently, a hyperparameter optimization algorithm, named grid search \cite{Bergstra12randomsearch}, was applied to the tuning of the optimal set of hyperparameters in the classification model, entitled support vector machine (SVM) \cite{708428}. For the SVM-based classification, the hyperparameters included kernel (linear, rbf, sigmoid), C (0.001, 0.01, 0.1, 1, 10, 25, 50, 100, 1000), and gamma (\enquote{auto}, 0.01, 0.001, 0.0001, 0.00001). By considering the grid search algorithm, the classification was implemented with a 10-fold stratified cross-validation. Finally, the prediction on the testing set of the classification model in each fold with optimal hyperparameters was evaluated. To compare the binary classification results within each MI/ME task for sit-to-stand and stand-to-sit transitions, a paired t-test with the unequal variances was used to determine the difference in the classification accuracy.

\begin{figure*}
\centering
\includegraphics[width=2\columnwidth]{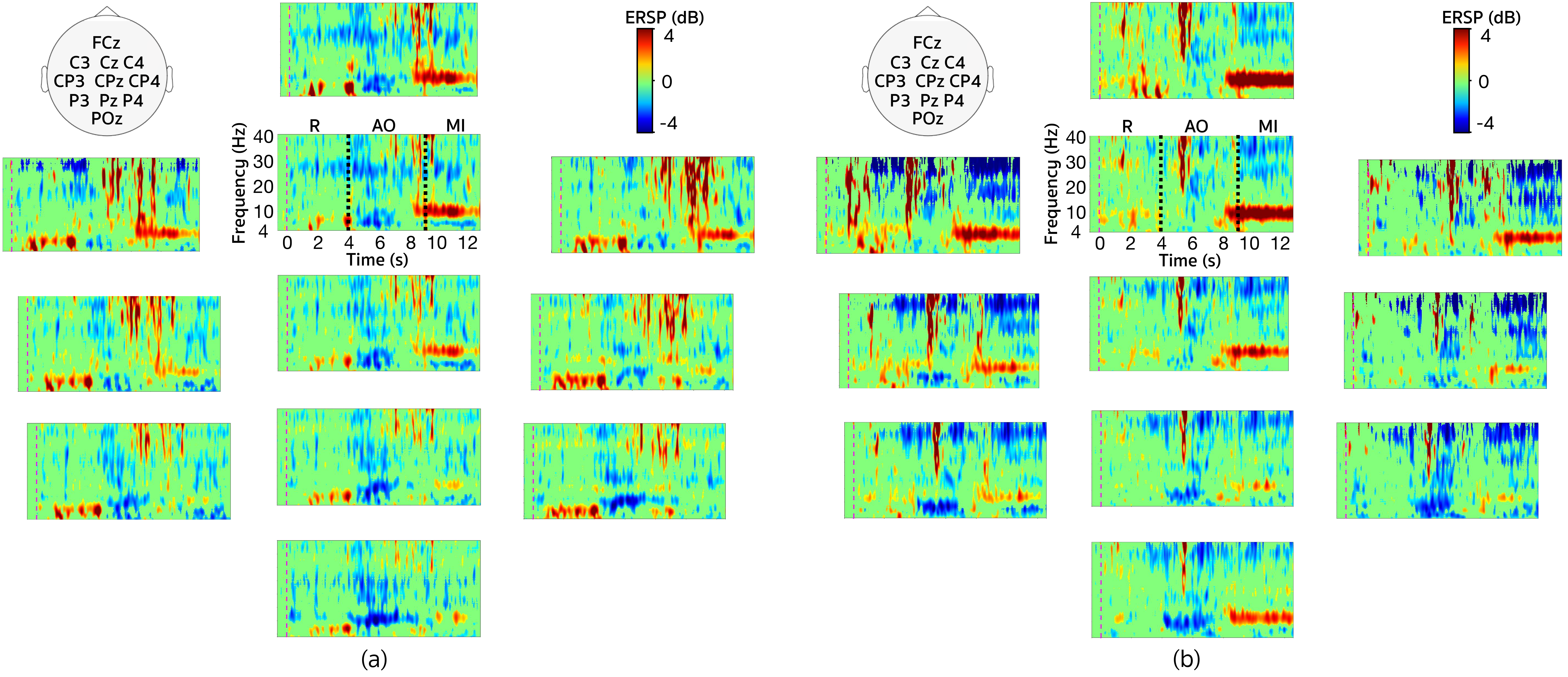}
\caption{Neural responses to MI sessions. Grouped event-related spectral perturbation (ERSP) for frequencies between 4--40 Hz across the entire trials were pooled for sit-to-stand (a) and stand-to-sit (b) tasks in comparison to the baseline of the R state (-1--0 s). The time interval from 0--4 s corresponds to the R state, 4--8 s corresponds to the AO state, 8--9 s corresponds to the idle state, and 9 s onward corresponds to the performing state. The sampling rate was set to 600 Hz for visualization. All present ERSP values were statistically significant compared to the baseline ($p=0.05$).}
\label{ersp_mi}
\end{figure*}

\begin{figure}
    \centering
    \includegraphics[width=0.99\columnwidth]{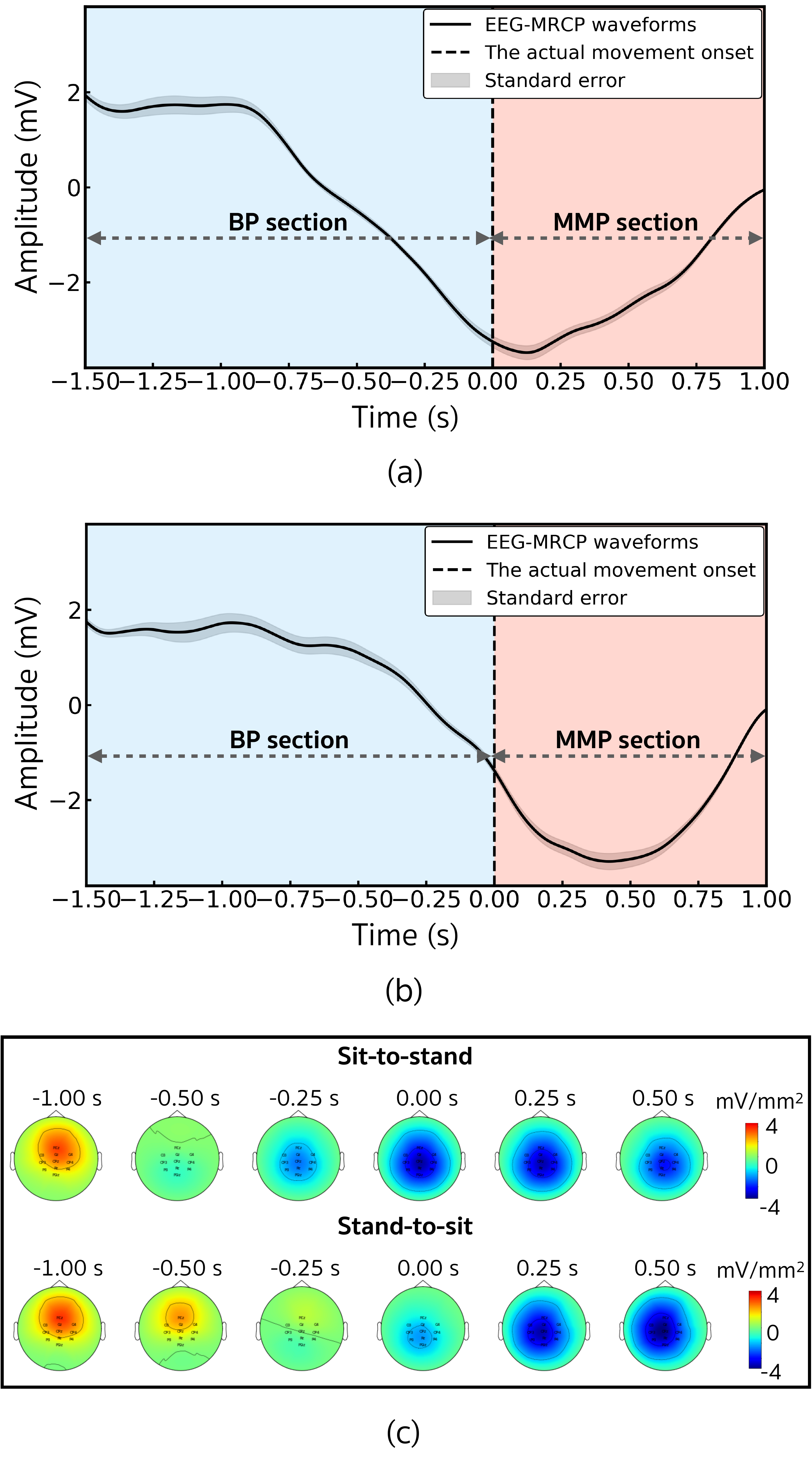}
    \caption{Grand average MRCP waveform (11 channels) across the 8 participants for the sit-to-stand (a) and stand-to-sit (b) transitions, from -1.5--1 s with respect to the actual movement onset. The scalp topographies (c) display the spatial representation of the change in MRCP amplitudes over time.}
    \label{mrcp_topo}
\end{figure}

\subsection{Classifier testing analysis}
Similar to the offline analysis, classifier testing analysis was performed using the LOOCV scheme with the same training models, the period of training set, and the grid search algorithm. Each epoch of the training set was likewise pre-processed with the same protocol as in the offline analysis to construct the continuous classifier testing model. For practical purposes, the period of the testing set was modified from 0--13 s for the MI session, whilst the duration was from 4--13 s in the ME session. By doing so, the testing set was streamed in segregated segments, each with a 2 s sliding window and a 0.2 s shift for the MI session. In order to investigate the feasibility of decoding the 3 classes of two step-binary classification models in the MI session (R versus AO and AO versus MI) for the sit-to-stand/stand-to-sit transitions, R versus AO classification model was used to evaluate the data in the first step. As shown in \autoref{pseudo_online_flowchart}, when the AO was produced after 5 consecutive detections ($AO\geq5$), the most optimal value was empirically selected among various participants. The algorithm was fashioned to determine which binary classification model was suitable. In the ME session, however, only AO versus MRCPs model was decoded, where the testing set was streamed in 1 s segment with a 0.5 s shift.

The performance of classifier testing analysis was calculated using 3 parameters: 

True positive rate (TPR) indicates the percentage of MI or MRCPs class, which was correctly decoded,
\begin{equation}
    \label{tpr}
    TPR = \frac{TP}{TP+FN}
\end{equation}

False positive rate (FPR) represents the percentage of MI or MRCPs class, which was detected during both R and AO states,
\begin{equation}
    \label{fpr}
    FPR = \frac{FP}{FP+TN}
\end{equation}

False negative rate (FNR) denotes the percentage of both R and AO classes, which could be detected during MI or ME states,
\begin{equation}
    \label{fnr}
    FNR = \frac{FN}{FN+TP}
\end{equation}

where $TP$ = True positive, $FP$ = False positive, $TN$ = True negative, and $FN$ = False negative.

To compare the binary classification results within each sit-to-stand/stand-to-sit-transition for MI and ME sessions, a paired t-test with unequal variances was used to determine the differences in the TPR, FPR, and FNR.

\section{Results}
This section aims to depict the findings and the key contributions amplified by each experiment. Result \textit{A.} reveals an investigated study of the MI signal and MRCP features; the ERSP and grand average MRCP waveforms are reported as the qualitative result. Result \textit{B.} leads us to the possibility of using MI signals and MRCP for BCI systems, which reveals the classification performance of decoding MI signals and MRCP in terms of offline and classifier testing analyses.

\subsection{Analysis of MI Signal and MRCPs Features} 
ERD/S have been studied widely in MI related works as one of the markers of brain responses. \autoref{ersp_mi} demonstrates the grouped ERSP across 8 participants in MI state from both sit-to-stand (left panel) and stand-to-sit (right panel) transitions. The ERSP delineates ERD/S from the entire duration of the trials with respect to the baseline spectra from 4 to 40 Hz. All present ERSP values were significant compared to the baseline ($p=0.05$). In comparison between the ERSP from 11 channels in all trials of both transitions, a significant decrease of alpha band power, indicating ERD, mainly in the parietal and parieto-occipital regions for the AO state (4--8 s) was found. However, a sustained increase of alpha band power, indicating ERS, for the performing MI (9 s onward) in fronto-central and central regions was observed. Furthermore, we observed an atypical increase of ERS in the performing state of stand-to-sit transition compared to sit-to-stand transition.

\autoref{mrcp_topo} illustrates the grand averages of the MRCPs (11 channels) across the 8 participants for the standing (1\textsuperscript{st} row) and sitting (2\textsuperscript{nd} row) movements. The MRCP waveform demonstrates a negative and a positive amplitude variation from -1.5 to 1 s with respect to the actual movement onset. Time 0 s was defined as the actual movement onset, in which the EMG signals overreached a pre-defined threshold amplitude. We observed the negative shape prior to the onset of actual movement (BP section), as well as the positive shape in MMP section. By considering the characteristic of the MMP section, we found a crucial difference between the amplitude pattern of the sit-to-stand and stand-to-sit transitions. There were appearances of the negative (the first 0.5 s period) and positive deflections (the last 0.5 s period) in the MMP section for the stand-to-sit transition. On the other hand, only the positive deflection in the MMP section of sit-to-stand transition was observed. The gray area along the MRCP waveforms denoted the SE of the BP and MMP amplitudes respective to each trial from the 8 participants. Scalp topographies (3\textsuperscript{rd} row) were plotted to display the spatial pattern distribution of the variation in the MRCP amplitude over time.

For MRCPs, the scalp distributions represented the average amplitude (11 channels) across all participants (120 trials) for the sit-to-stand and stand-to-sit actions. Based on the topographies, we observed the brain activities during the intention of movement by dividing the time interval into two phases. Prior to the beginning of the sitting and standing movements or during motor planning (-1.5 to 0 s), the brain activity displayed a slow-rising negative distribution in the central brain areas. After actual movement onset in ME session (0 to 1 s), the power of the spatial pattern returned from negative to positive spatial pattern.

\begin{figure*}[t]
    \centering
    \includegraphics[width=1.95\columnwidth]{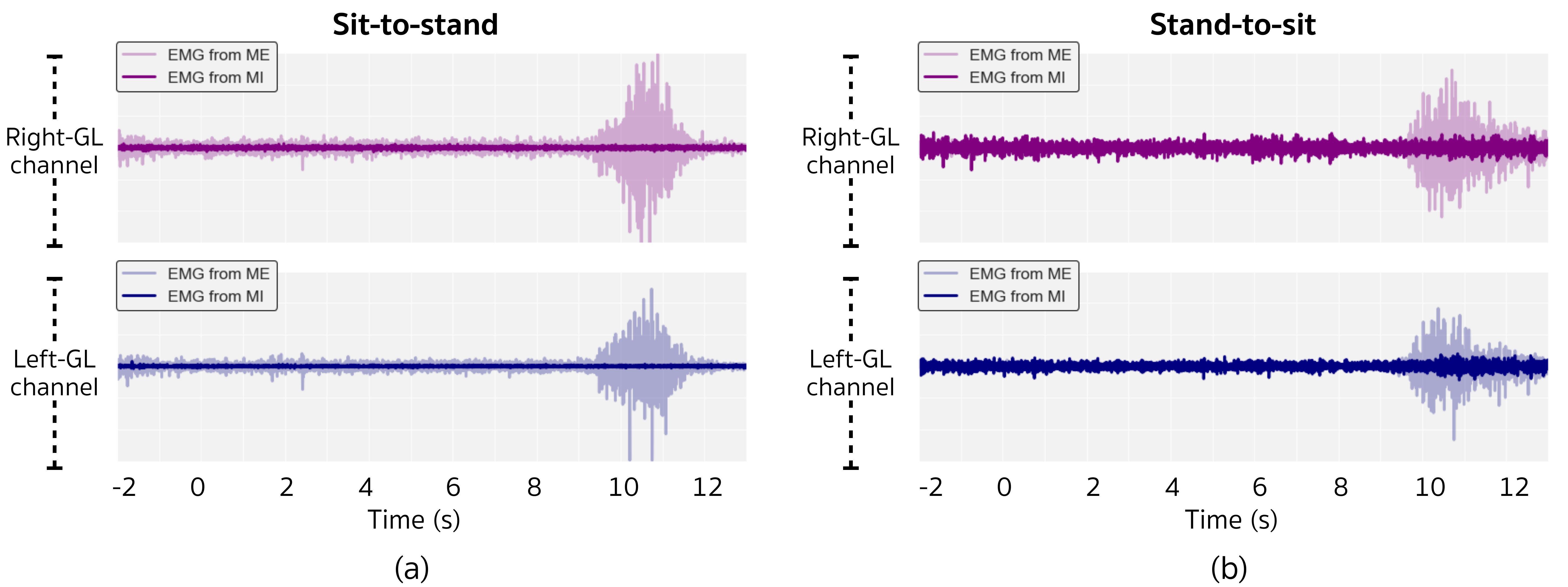}
     \caption{Grand average filtered EMG waveform across 8 participants from right and the gastrocnemius lateralis (GL) channels for sit-to-stand (a) and stand-to-sit (b) transitions.}
    \label{compare_EMG}
\end{figure*}

\autoref{compare_EMG} indicates the grand averages of the filtered EMG signals (filtered between 15--124 Hz) across all participants (120 trials) for both MI and ME sessions. There were no any actual movements on the MI compared to the ME for both the sit-to-stand \autoref{compare_EMG}(a) and stand-to-sit \autoref{compare_EMG}(b) transitions.
\newpage

\begin{table}[]
    \caption{Classification accuracy (in \%) of sit-to-stand (sit:std) and stand-to-sit(std:sit) transitions during the MI and ME sessions. Bold and * denote that the number is significantly higher than the other movements, $p < 0.05$}
    \label{ERD/S_and_MRCPs_results}
    \centering
    \resizebox{1\columnwidth}{!}{%
        \begin{tabular}{@{}ccccccc@{}}
        \toprule[0.2em]
        \multirow{3}{*}{\textbf{Subject ID}} & \multicolumn{4}{c}{\textbf{MI Session}}                                        & \multicolumn{2}{c}{\textbf{ME Session}} \\ \cmidrule[0.1em](l){2-5}\cmidrule[0.1em](l){6-7} 
                    & \multicolumn{2}{c}{\textbf{R vs. AO}} & \multicolumn{2}{c}{\textbf{AO vs. MI}}  & \multicolumn{2}{c}{\textbf{AO vs. MRCPs}} \\ \cmidrule[0.1em](l){2-3}\cmidrule[0.1em](l){4-5} \cmidrule[0.1em](l){6-7} 
                                            & \textbf{sit:std} & \textbf{std:sit} & \textbf{sit:std} & \textbf{std:sit} & \textbf{sit:std} & \textbf{std:sit} \\ \midrule[0.1em]
        \textbf{S1} & 71.82         & 84.24 & 85.76 & 95.45         & 85.83                   & 88.33           \\ 
        \textbf{S2} & 60.61         & 46.97 & 71.82 & 80.61         & 65.00                   & 72.50           \\ 
        \textbf{S3} & 47.58         & 72.42 & 67.88 & 79.70         & 68.33                   & 65.83           \\ 
        \textbf{S4} & 66.36         & 59.09 & 61.52 & 82.73         & 75.00                   & 66.67           \\ 
        \textbf{S5} & 63.94         & 57.58 & 79.70 & 86.36         & 77.50                   & 65.83           \\ 
        \textbf{S6} & 66.67         & 63.64 & 73.64 & 69.70         & 70.83                   & 65.83           \\ 
        \textbf{S7} & 68.48         & 70.61 & 85.76 & 83.03         & 93.33                   & 72.50           \\ 
        \textbf{S8} & 72.73         & 55.76 & 83.03 & 84.24         & 77.50                   & 88.33           \\ \midrule[0.1em] 
        \textbf{Mean}                        & 64.77    & 63.79            & 76.14             & \textbf{82.73*}   & 76.67    & 73.23            \\ 
        \textbf{$\pm$SE} & $\pm$2.82 & $\pm$4.11  & $\pm$3.14  & \textbf{$\pm$2.54} & $\pm$3.29           & $\pm$3.44            \\ \bottomrule[0.2em]
        \end{tabular}%
    }
\end{table}

\begin{table*}[]
\caption{TPR, FNR, and FPR results (in \%) from MI and ME sessions with personalized classifier testing analysis. Bold and * represents the number which was significantly higher than the other tasks, $p < 0.05$}
     \label{Result_Pseudo-online_part}
     \centering
     \resizebox{1.9\columnwidth}{!}{
     \begin{tabular}{@{}ccccccccccccc@{}}
        \toprule[0.2em]
        \multirow{3}{*}{\textbf{Subject ID}} & 
        \multicolumn{6}{c}{\textbf{Sit-to-stand}} & \multicolumn{6}{c}{\textbf{Stand-to-sit}} \\ \cmidrule[0.1em](l){2-7} \cmidrule[0.1em](l){8-13}
         & \multicolumn{2}{c}{\textbf{TPR}} & \multicolumn{2}{c}{\textbf{FPR}} & \multicolumn{2}{c}{\textbf{FNR}} & \multicolumn{2}{c}{\textbf{TPR}} & 
         \multicolumn{2}{c}{\textbf{FPR}} & \multicolumn{2}{c}{\textbf{FNR}} \\  
         \cmidrule[0.1em](l){2-3} \cmidrule[0.1em](l){4-5}  \cmidrule[0.1em](l){6-7}  \cmidrule[0.1em](l){8-9}  \cmidrule[0.1em](l){10-11}  \cmidrule[0.1em](l){12-13} 
         & \textbf{MI} & \textbf{ME} & \textbf{MI} & \textbf{ME} & \textbf{MI} & \textbf{ME} & \textbf{MI} & \textbf{ME} & \textbf{MI} & \textbf{ME} & \textbf{MI } & \textbf{ME} \\ \midrule[0.1em]
        \textbf{S1} & 69.63 & 62.89 & 7.54 & 37.22 & 30.37 & 37.11 & 79.26 & 85.33 & 8.60 & 23.44 & 20.74 & 14.67 \\
\textbf{S2} & 60.00 & 50.67 & 18.25 & 38.10 & 40.00 & 49.33 & 48.52 & 70.67 & 14.04 & 47.62 & 51.48 & 29.33 \\
\textbf{S3} & 62.59 & 64.00 & 19.12 & 52.86 & 37.41 & 36.00 & 78.89 & 72.00 & 23.16 & 68.57 & 21.11 & 28.00 \\
\textbf{S4} & 54.07 & 66.67 & 20.70 & 40.95 & 45.93 & 33.33 & 70.74 & 64.00 & 15.09 & 50.00 & 29.26 & 36.00 \\
\textbf{S5} & 57.04 & 78.67 & 10.00 & 44.76 & 42.96 & 21.33 & 71.48 & 80.00 & 10.35 & 49.05 & 28.52 & 20.00 \\
\textbf{S6} & 57.04 & 66.67 & 17.72 & 43.33 & 42.96 & 33.33 & 55.56 & 68.44 & 20.70 & 61.68 & 44.44 & 31.56 \\
\textbf{S7} & 78.89 & 69.33 & 16.49 & 41.43 & 21.11 & 30.67 & 80.37 & 68.89 & 17.37 & 64.98 & 19.63 & 31.11 \\
\textbf{S8} & 84.07 & 66.67 & 14.39 & 42.86 & 15.93 & 33.33 & 82.22 & 69.33 & 20.70 & 46.19 & 17.78 & 30.67 \\\midrule[0.1em] 
\textbf{Mean} & 65.42 & 65.70 & 15.53 & \textbf{42.69*} & 34.58 & 34.30 & 70.88 & 72.33 & 16.25 & \textbf{51.44*} & 29.12 & 27.67 \\
\textbf{$\pm$SE} & $\pm$3.90 & $\pm$2.74 & $\pm$1.63 & \textbf{$\pm$1.71} & $\pm$3.90 & $\pm$2.74 & $\pm$4.41 & $\pm$2.45 & $\pm$1.83 & \textbf{$\pm$5.03} & $\pm$4.41 & $\pm$2.45 \\ \bottomrule[0.2em]
        \end{tabular}
     }
\end{table*}

\subsection{Classification Performance of Decoding MI Signal and MRCPs}

The average performance across all participants of the proposed MI (R versus AO and AO versus MI) and ME (AO versus MRCPs) for binary classification are displayed in \autoref{ERD/S_and_MRCPs_results}. The classification performance comparison between the sit-to-stand and the stand-to-sit transitions was used for MI signals and MRCPs decoding. The result of the MI (R versus AO and AO versus MI) revealed that the action of stand-to-sit transition significantly outperformed the sit-to-stand transition throughout the binary classification between AO and MI, $t(231) = -2.54$, $p < 0.05$, whereas the classification of R and AO did not exhibit significant difference. 

In the ME session, according to the data obtained from the participants performing, no MRCP was detected during R and AO states in the timeline of the experimental trials, whereas the MRCPs were detected during the task performing state. Hence, it was decided to perform only the classification task of AO versus MRCPs. The result of ME (AO versus MRCPs) failed to provide a statistically significant difference between these two transitions.

\begin{figure*}[t]
\centering
\includegraphics[width=1.9\columnwidth]{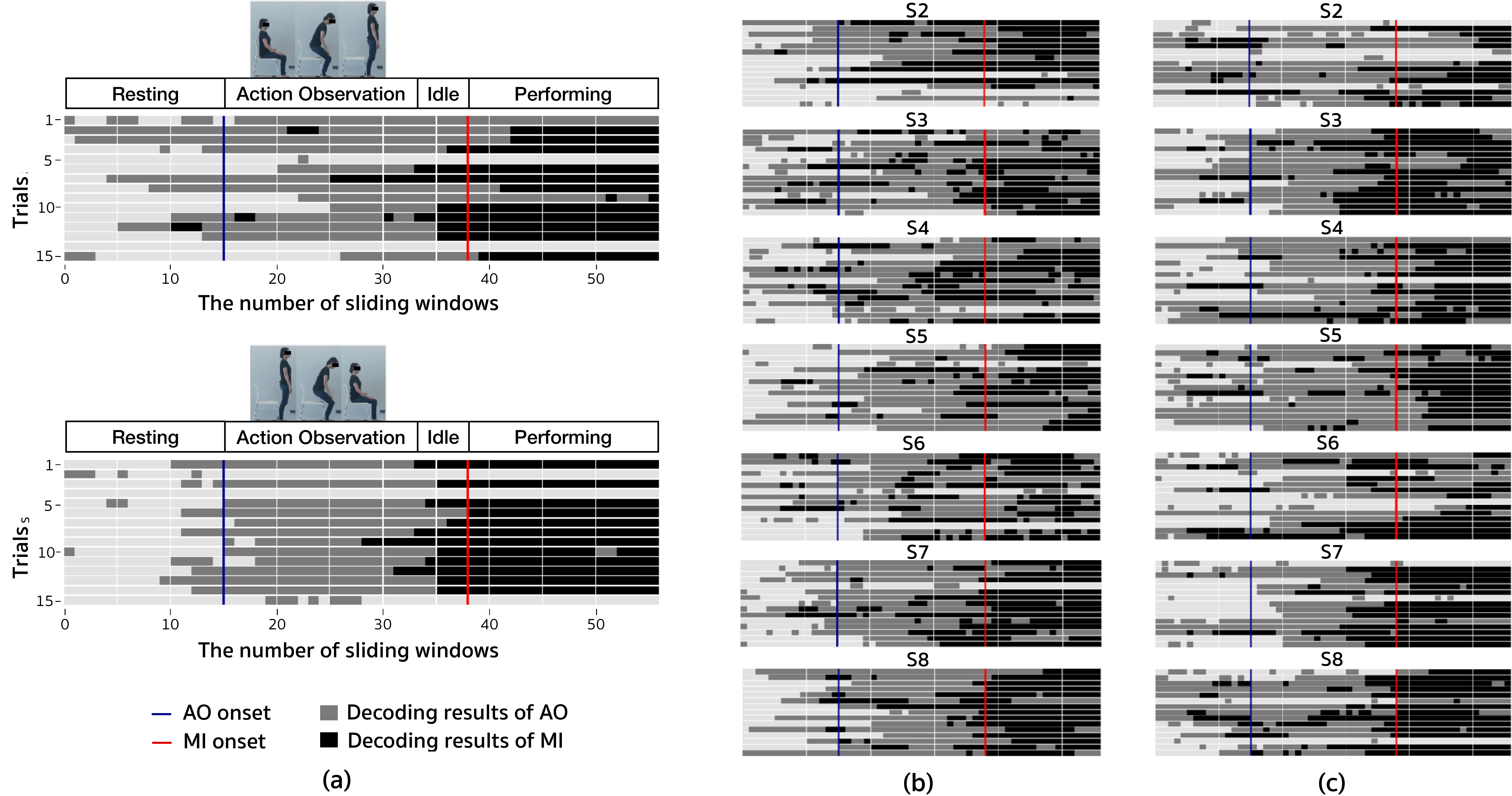}
\caption{Representation of the personalized classifier testing output of MI tasks. (a) illustrates the decoding output from sit-to-stand transition (top panel) and stand-to-sit (bottom panel) transition of one participant, while (b) and (c) demonstrate the decoding result of sit-to-stand and stand-to-sit transitions of the other participants respectively. The light grey, dark grey, and black squares indicate the decoding output for the R, AO, and MI classes respectively.}
\label{Pseudo-online}
\end{figure*}

The TPR, FPR, and FNR results of classifier testing analysis are displayed in \autoref{Result_Pseudo-online_part}. The decoding results of the MI (R versus AO and AO versus MI) and ME (AO versus MRCPs) during the sit-to-stand transitions and stand-to-sit transitions were compared. The grand average FPR or \textit{false alarm rate} from both transitions demonstrated that the coherent results from the ME are significantly higher than the MI, with $t(238) = -12.62$, $p < 0.05$ for the sit-to-stand transitions and $t(230) = -13.94$, $p < 0.05$ for stand-to-sit transitions. On the other hand, the grand average TPR and FNR results did not indicate significant differences between these two sessions. The comparison of the grand average FPR, TPR and FNR are further discussed in \textit{Discussion C}. In addition, \autoref{Pseudo-online} illustrates the representation of the classifier testing results of MI (R versus AO and AO versus MI). The total number of windows throughout each trial was 56. The blue line indicates the AO onset, while the red line refers to MI onset. The light grey, dark grey, and black boxed indicate the decoding results for the R, AO, and MI classes respectively. Each window was decoded from the combined binary classification model between R versus AO and AO versus MI.

\section{Discussion}

\subsection{Characteristics of ERD/S During Action Observation and Motor Imagery}
The current study investigated the role of action observation (AO) and motor imagery (MI) during the standing and sitting tasks. Specifically, the EEG potentials during resting (R) and the task performance (MI/ME) of sit-to-stand and stand-to-sit transitions were examined. Here, we introduced video presentations during the experiments for the simplification of the instruction, which distinguished the current study from the previous works. In our case, the videos of an actual person performing the acts of sit-to-stand and stand-to-sit were shown to the participants prior to task performing (MI/ME) state. On the other hand, in other similar studies, participants were simply shown a few words or symbols, as visual cues, without being explicitly told about the particular actions that were instructed to perform. This could be the reason why ERD was observed during AO and ERS was observed during MI tasks in our study, while ERD/S was only observed during MI in the previous studies \cite{cho2017eeg, lee2019eeg}. However, another study demonstrated greater ERD/S power during AO in comparison to MI, which supports our findings in terms of usefulness of the future treatments for patients who have limited MI ability \cite{tani2018action}. 

Although we did not take into account the perspective-dependent action in our study, there were studies showing the effect of EEG rhythms from viewed self-performance \cite{nagai2019action, song2019paradigm}. Specifically, the ERD/S response for observing a participant's own hand was stronger than when the participants observed the movement of another person's hand. Future research may take account of the perspectives in the design of their study, where participants can be asked to view the motor actions executed by themselves or by another person.

\subsection{Decoding algorithms for standing and sitting tasks}

The current study aimed to compare and differentiate the EEG rhythms during the sit-to-stand and stand-to-sit transitions. According to the AO versus MI in \autoref{ERD/S_and_MRCPs_results}, the mean accuracy was highest for the stand-to-sit transition at 82.73$\pm$2.54\%, which was statistically higher than the sit-to-stand transition. This suggests that the MI activation during the sit-to-stand transition was distinguishable from the stand-to-sit transition, corresponding to the characteristics of grand average MRCPs shown in \autoref{mrcp_topo}. The latency in the MRCPs of the stand-to-sit transition reflected the more difficult nature of this transition (i.e., the lack of visual feedback towards the back as a person moved from standing to sitting) in comparison to the sit-to-stand transition, making it easier for the classifier to distinguish between AO and MI in this former transition. Indeed, previous studies have shown the effects of task complexity on ERD/S rhythms \cite{qiu2017optimized,mashat2019effects}. To overcome the limitations of the current study, higher number of participants as well as number of trials are required. Consequently, there is a possibility to apply our developed deep learning approaches to increase the accuracy of the classification of EEG rhythms in the future \cite{wilaiprasitporn2019affective, ditthapron2019universal}.

\subsection{Feasibility study and applications}

The final aspect of the current study is to apply our findings to the future development of BCI-based exoskeleton system used for the rehabilitation of patients with motor disorders. Currently, no sufficient decoding performance has been obtained from the MI-BCI systems. The usability of the MI task has often increased the user’s fatigue. It is also difficult to predict the exact movement from human imagination. On the other hand, ME task has been shown to provide a higher decoding performance compared to MI task \cite{9052997}. In this study, we performed the classifier testing scheme on our continuous EEG rhythms across R, AO, and MI/ME. TPR, FPRs, and FNR variables were observed and were used to statistically control for any differences across sessions. As reported in \autoref{Result_Pseudo-online_part}, the comparison of the MI (R versus AO and AO versus MI) and ME (AO versus MRCPs) has demonstrated that the TPR and FNR did not display any statistical difference. However, the FPRs or \textit{false alarm rates} of sit-to-stand and stand-to-sit transitions during the MI session were significantly lower than those in the ME session.

This causal effect revealed an interesting fact about the desired movements, which were performed when the TPR was high and FPRs were low, in contrast to the unintended performed movements which led to an increase in FPR. Thus, the EEG rhythms from AO and MI states were more feasible than those from the ME session due to the ability to decode the desired movements by deliberately maintaining the high level of TPR and the low level of FPRs. The experimental protocol from the present study will be suitable for BCI-based exoskeleton systems for rehabilitating patients with motor disorders of the lower limbs in future studies. This has been supported by the previous studies as the signals during MI have been found to facilitate the operation of lower limb exoskeleton when the classification accuracy of MI patterns, and even the motor intention, was high \cite{vinoj2019brain, delisle2019system}.

\section{Conclusion}
In this paper, we investigated the possibility of combining action observation (AO) and motor imagery (MI) as one of the aspects for a brain-computer interface (BCI) system (e.g., exoskeleton-based rehabilitation). We created a behavioral task in which the participants were instructed to perform both AO and MI/motor execution (ME) in regard to the actions of sit-to-stand and stand-to-sit. The pattern discrimination revealed that ERD responded during AO while ERS responded during MI at the alpha band across the sensorimotor area. We obtained promising experimental results from both offline and classifier testing analyses by using leave-one(trial)-out cross-validation (LOOCV) scheme. The integration of the filter bank common spatial pattern (FBCSP) and support vector machine (SVM) performed well in decoding the neural intentions between AO and MI for both offline and classifier testing analyses. Together, our results suggest the feasibility of using the future exoskeleton-based rehabilitation that combines both AO and MI.   

\bibliography{References}
\bibliographystyle{IEEEtran}
\end{document}